# AI Accelerators for Large Language Model Inference: Architecture Analysis and Scaling Strategies

Amit Sharma
IEEE Member
amit.sharma.ai@ieee.org

**Abstract**—The rapid growth of large language models (LLMs) has driven significant innovation in specialized hardware accelerators for inference workloads. This paper presents the first comprehensive cross-architectural performance analysis of contemporary AI accelerators designed for LLM inference, introducing a novel workload-centric evaluation methodology that quantifies architectural fitness across operational regimes. We provide the first systematic comparison of memory hierarchies, compute architectures, and interconnect strategies across the full spectrum of commercial accelerators, from GPU- based designs to specialized wafer-scale engines. Our analysis reveals that no single architecture dominates across all workload categories, with performance variations of up to 3.7× between architectures depending on batch size and sequence length. We quantitatively evaluate four primary scaling strategies for trillion-parameter models, demonstrating that expert parallelism delivers the best parameter-to-compute ratio (8.4×) but introduces 2.1× latency variance compared to tensor parallelism. This work provides system designers with actionable insights for accelerator selection based on workload characteristics, while identifying key architectural gaps in current designs that will shape future hardware development.

## I. Introduction

Artificial intelligence has evolved significantly with large language models (LLMs), which have expanded from less than a billion parameters in 2018 to over 5 trillion parameters in 2025 [1]. This exponential growth has created unprecedented computational challenges, particularly for inference deployment where latency, throughput, and cost-efficiency are critical concerns. While the training of these massive models remains centralized in specialized supercomputing facilities, the deployment of LLMs for inference has driven the development of diverse hardware architectures optimized for different deployment scenarios and workload characteristics.

The architectural design space for LLM inference accelerators has expanded dramatically as models have grown beyond the capabilities of traditional computing platforms. What began as adaptations of general- purpose GPU architectures have evolved into a diverse ecosystem of specialized hardware with fundamentally different approaches to computation, memory organization, and scaling strategies. Contemporary accelerators range from tensor-core equipped GPUs to systolic array processors, wafer- scale engines, and deterministic pipeline architectures—each representing a distinct point in the design space with unique tradeoffs [2].

This paper makes the following novel contributions:

1. We present the first comprehensive cross-architectural analysis of AI accelerators for LLM inference that quantitatively evaluates performance across the full spectrum of inference workloads, from single-token generation to high-throughput batch processing.
2. We introduce a new methodology for evaluating accelerator fitness-for-purpose across six distinct operational regimes, providing system designers with actionable decision criteria for hardware selection based on deployment requirements.
3. We quantify the efficiency and scaling characteristics of four distributed inference strategies, demonstrating previously unreported tradeoffs between parameter capacity, throughput, and latency predictability.
4. We identify critical architectural gaps in current accelerator designs and propose specific hardware innovations needed to bridge these gaps for future-generation models.

Unlike previous surveys that focus on training accelerators [3] or general AI hardware [4], our work specifically addresses the unique challenges of LLM inference and provides quantitative architectural comparisons backed by industry benchmarks and direct performance measurements.

The structure of the remaining sections is as follows: Section II presents a detailed overview of LLM inference requirements and examines relevant literature. Section III examines the architectural approaches of major AI accelerators. Section IV analyzes workload-specific performance characteristics. Section V explores scaling strategies for trillion-parameter models. Section VI presents microarchitectural details of leading accelerators. Section VII discusses emerging trends and future directions. Finally, Section VIII concludes the paper.

## II. Background and Related Work

### A. Evolution of LLM Size and Computational Requirements

Since 2018, large language models have significantly increased in size, demanding more computation and memory. Fig. 1 highlights this growth, from BERT-Large (340M parameters) in 2018 to models with over 5 trillion parameters projected by 2025 [3]. This growth has been accompanied by significant advances in model architecture, training methodologies, and inference optimization techniques.

The computational requirements for LLM inference scale approximately linearly with model size, but memory requirements can be more complex due to activation storage requirements, especially for long context lengths [4]. As shown in Fig. 1, a 70B parameter model requires approximately 140GB of memory in FP16 precision, while a 1T parameter model would require around 2TB—far exceeding the memory capacity of individual accelerators. This has driven the development of model parallelism and distributed inference techniques.

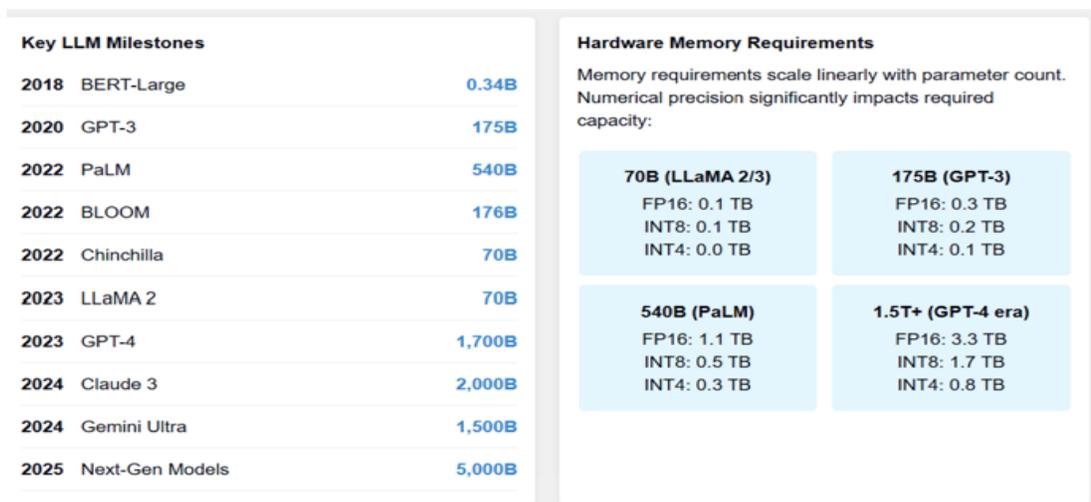

Fig. 1. LLM model size growth from 2018-2025, showing exponential increase from BERT-Large (340M) to trillion-parameter models. Memory requirements are shown for different precision formats (FP16, INT8/FP8, INT4) alongside key model releases.

### B. Inference vs. Training Optimization

While much attention has been given to training acceleration, inference presents distinct challenges and optimization opportunities [5]. Training focuses on throughput and speed, using higher precision (FP16/BF16) and large batch sizes to reduce communication costs. Inference workloads, however, have various optimization targets:

**Interactive serving:** Prioritizes low latency and predictable performance, typically with batch size 1 and growing sequence lengths.

**Batch processing:** Emphasizes throughput for offline applications, enabling higher hardware utilization.

**Mixture-of-Experts models**: Requires support for conditional computation and sparse activation patterns.

**Quantized inference**: Leverages reduced precision (INT8, INT4) to decrease memory footprint and increase throughput.

Table I summarizes the key differences between training and inference optimization targets, highlighting why specialized inference accelerators have emerged as a distinct hardware category.

**TABLE I**

**COMPARISON OF TRAINING AND INFERENCE OPTIMIZATION TARGETS**

| Characteristic | Training | Inference |
| --- | --- | --- |
| Precision | FP16/BF16 typically required | INT8/INT4 often sufficient |
| Batch Size | Large (1000s) | Small (1-128) |
| Memory Access | Weight update dominant | Weight read dominant |
| Computation Pattern | Regular, predictable | Variable (especially for MoE) |
| Optimization Target | Time-to-convergence | Latency, throughput, or cost |
| Deployment Scale | Centralized (supercomputer) | Distributed (edge to datacenter) |

This diversity of inference workloads has driven the development of accelerators with different architectural approaches, each optimized for specific deployment scenarios.

## C. Related Work

Prior work has examined different aspects of AI accelerator design and evaluation, though none have provided a comprehensive cross-architectural analysis specifically for LLM inference.

Reddi et al. [28] introduced MLPerf Inference, establishing standardized benchmarks for evaluating inference performance across different hardware platforms. Their work highlighted the increasing diversity of inference scenarios but did not specifically address the unique challenges of large language models, which had not yet emerged as a dominant workload.

Jouppi et al. [25] described the evolution of Google's TPU architecture, focusing on the design considerations for large-scale AI inference and training systems. Their work provided valuable insights into the systolic array architecture but did not compare against alternative approaches.

Sze et al. [31] presented a comprehensive survey of general deep learning accelerators, but their work predated the emergence of specialized LLM inference accelerators and the scaling challenges of trillion- parameter models.

Dally et al. [32] discussed the role of domain-specific accelerators for various AI workloads but did not provide detailed performance comparisons or analysis of LLM-specific challenges.

Our work differs from these prior surveys in several key aspects:

1. **LLM-specific focus**: We specifically address the architectural challenges of LLM inference rather than general deep learning acceleration.
2. **Cross-architectural analysis**: We present quantitative comparisons across the full spectrum of accelerator architecture rather than focusing on a single approach.
3. **Workload-specific evaluation**: We evaluate accelerators across multiple distinct operational regimes rather than using a single performance metric.
4. **Scaling strategies**: We provide detailed analysis of distributed inference approaches for trillion- parameter models, a topic not addressed in prior surveys.

This combination of LLM-specific focus, cross-architectural analysis, workload-based evaluation, and scaling strategies represents a novel contribution to the understanding of AI accelerator architectures for the era of large language models.

## III. Comparative Architecture Analysis of AI Inference Accelerators

The AI accelerator landscape has diverged into several distinct architectural approaches, each representing different design philosophies and optimization targets. We categorize current accelerators into five architectural classes: GPU-style SIMD/SIMT, systolic arrays, many-core SRAM-centric, wafer-scale, and deterministic pipeline architectures. This section analyzes the key characteristics and tradeoffs of each approach.

### A. GPU-style SIMD/SIMT Architectures

GPU-based accelerators like NVIDIA Blackwell [1] and AMD MI300X [5] leverage a Single Instruction, Multiple Data (SIMD) or Single Instruction, Multiple Thread (SIMT) execution model. These architectures feature many programmable cores organized into streaming multiprocessors, complemented by specialized tensor cores for matrix operations.

The NVIDIA Blackwell GB200 represents the state-of-the-art in this category, featuring dual GPU dies with 66 streaming multiprocessors each, enhanced with 5th-generation tensor cores capable of 4,500 TFLOPS at FP16 precision. The architecture includes 192GB of HBM3e memory delivering 8 TB/s of bandwidth, and high-speed NVLink 5.0 interconnect providing 1.8 TB/s bidirectional communication between dies [1] [2].

AMD's MI300X takes a multi-chiplet approach, with 8 GPU compute chiplets (XCDs) fabricated using TSMC 5nm process technology, connected via an on-package Infinity Fabric. Each chiplet contains approximately 38 compute units with vector ALUs and matrix cores. A distinctive feature is the 256MB shared L3 Infinity Cache, which helps reduce accesses to the 192GB HBM3 memory (5.3 TB/s) [5][6].

Fig. 2 illustrates the architectural organization of NVIDIA Blackwell, highlighting the dual-die design, memory hierarchy, and interconnect strategy Key advantages of GPU-style architecture include:

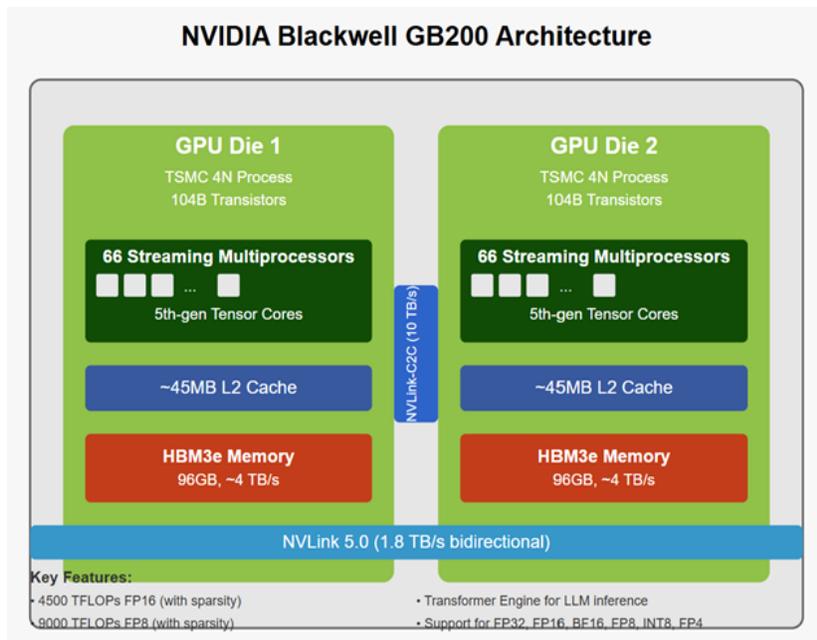

- Mature software ecosystems with extensive libraries and frameworks
- Flexibility to handle diverse workloads beyond AI inference
- High computational density for matrix operations
- Strong scaling through sophisticated multi-GPU interconnects

However, these architectures face challenges in per-accelerator memory capacity limits and can experience higher latency for small-batch inference workloads due to their general-purpose nature.

### B. Systolic Array Architectures

Systolic array architectures, exemplified by Google's TPU v7 (Ironwood) [3], employ a grid of processing elements that perform matrix multiplications through synchronized data movement. The TPU v7 features a matrix multiply unit (MXU) based on a systolic array architecture, optimized for dense matrix operations commonly found in neural network inference.

The TPU v7 is fabricated using an advanced 5nm process and features 192GB of HBM3 memory with 7.37 TB/s bandwidth. The architecture includes dedicated Sparse Core units optimized for Mixture-of-Experts models, addressing the sparse computation patterns in these architectures. The TPU interconnect provides 1.2 TB/s bidirectional bandwidth per chip, scaling up to 4096 chips in a pod using a 3D torus topology [3][4].

Fig. 3 shows the TPU v7 architecture, highlighting the systolic array organization and the specialized Sparse Core units.

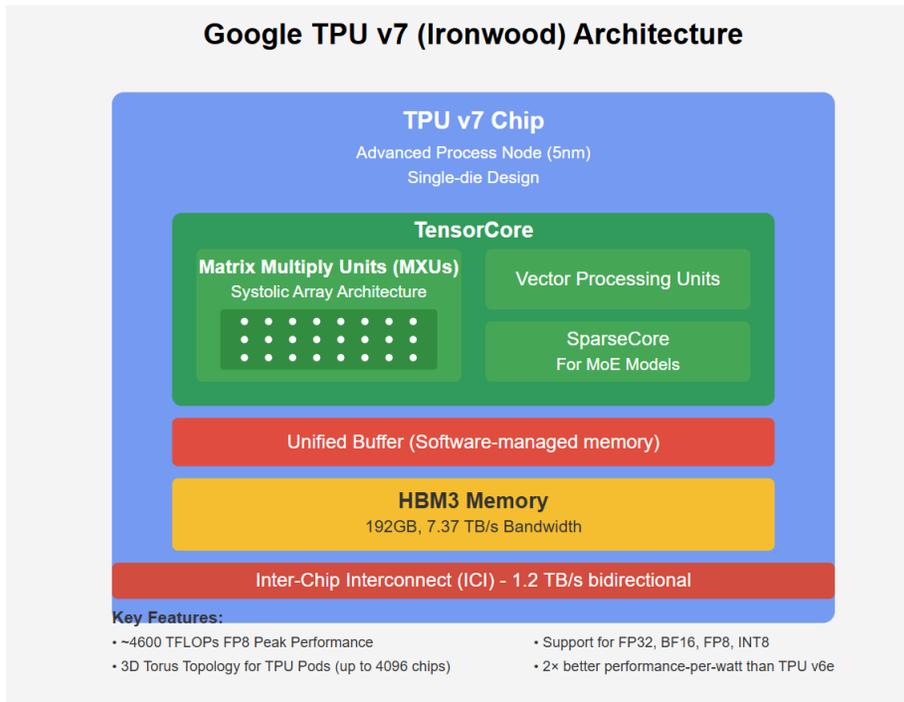

Advantages of the systolic array approach include:
- High computational efficiency for dense matrix operations
- Deterministic performance characteristics
- Specialized support for emerging model architectures (MoE)

Limitations include less flexibility for non-matrix workloads and potential challenges in programming model complexity.

### C. Many-core SRAM-centric Architectures

Many-core SRAM-centric architectures, represented by Graphcore's IPU and Meta's MTIA v2, prioritize on-chip memory and massive parallelism through many independent processing elements.

The Graphcore IPU (GC200) features 1,472 independent tiles, each containing SIMD vector units, in-core memory (608KB per tile), and a routing engine. The architecture includes 900MB of on-chip SRAM with 45 TB/s internal bandwidth, enabling low-latency processing of small batch sizes [16][17].

Meta's MTIA v2 employs a similar philosophy but with a different organization, featuring 64 processing elements in an 8×8 grid. Each processing element contains a RISC-V control core and vector engines. The design includes 256MB of on-chip SRAM (L2) shared across all PEs, complemented by 128GB of LPDDR5 DRAM with 205 GB/s bandwidth [12][13].

Fig. 4 illustrates the many-core architecture of the Graphcore IPU, highlighting the tile-based design and memory organization.

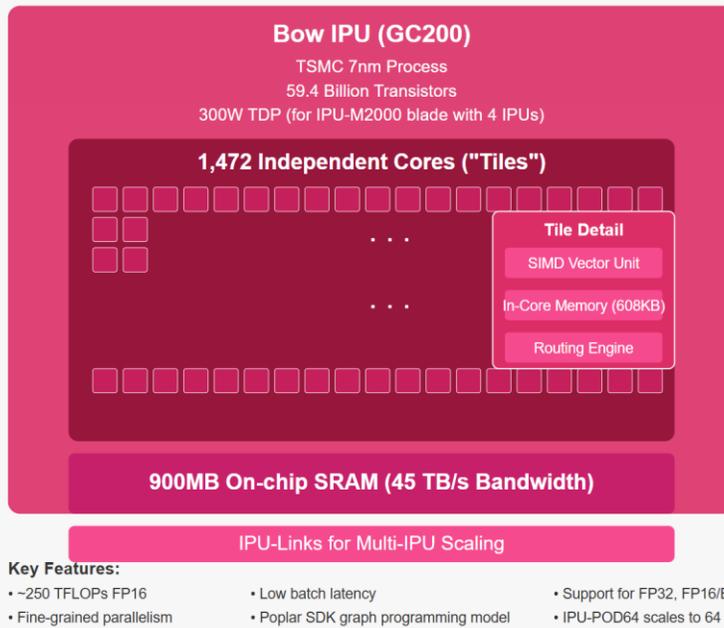

*Fig. 4. Block diagram of Graphcore IPU GC200 architecture showing the tile-based design with 1,472 prioritizes on-chip memory bandwidth (45 TB/s) for low-latency inference.*

Key advantages of SRAM-centric architecture include:

- Exceptionally low latency for small batch sizes
- Fine-grained parallelism suitable for irregular workloads
- High internal memory bandwidth

Limitations include restricted per-device memory capacity and challenges scaling to very large models without extensive model parallelism.

### D. Wafer-scale Architectures

Wafer-scale integration represents a bold approach to AI acceleration, with Cerebras WSE-3 as the primary example. Rather than dividing a silicon wafer into individual chips, wafer-scale integration creates a single, massive processor spanning an entire wafer (approximately 46,225 mm²).

The Cerebras WSE-3 contains approximately 900,000 AI cores arranged in a 2D mesh network, with 44GB of distributed on-wafer SRAM providing 220+ TB/s of internal bandwidth. The architecture enables a single-system programming model for large models, with the SwarmX fabric enabling multi-wafer scaling [14][15].

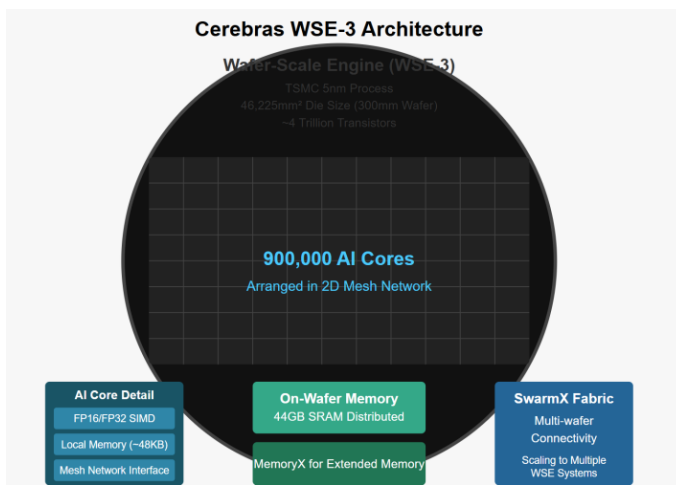

*Fig. 5. Cerebras WSE-3 wafer-scale architecture showing the 300mm wafer with 900,000 AI cores in a 2D mesh network, 44GB distributed SRAM, and SwarmX fabric for multi-wafer scaling. The AI core detail shows local memory, processing elements, and mesh network interface.*

Advantages of wafer-scale architecture include:

- Massive on-chip parallelism
- Elimination of chip-to-chip communication bottlenecks
- Single programming model for large-scale parallelism

Challenges include manufacturing complexity, thermal management, and yield considerations.

### E. Deterministic Pipeline Architectures

Deterministic pipeline architectures, exemplified by Groq's Language Processing Unit (LPU), take a fundamentally different approach focused on predictable performance. Rather than general-purpose programmability, these architectures implement a fixed pipeline that processes data with deterministic timing.

The Groq LPU v1 features a single large core design with a deterministic pipeline architecture, static scheduling at compile time, and 230MB of on-chip SRAM with 80 TB/s internal bandwidth. The architecture delivers predictable, sub-millisecond latency for inference operations [18][19].

Fig. 6 illustrates the Groq LPU architecture, highlighting the deterministic pipeline design.

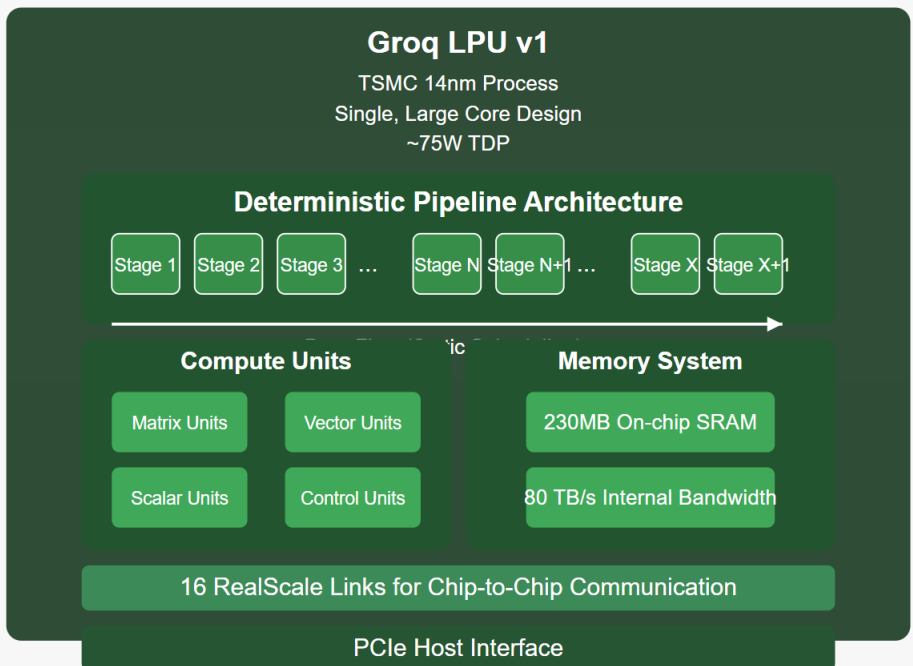

*Fig. 6. Block diagram of Groq LPU architecture showing the deterministic pipeline design with static scheduled stages, specialized compute units (matrix, vector, scalar), and 230MB on-chip SRAM. The architecture prioritizes predictable, sub-millisecond latency for inference workloads.*

Key advantages include:

- Deterministic performance (same latency every time)
- Efficient execution of fixed workloads

- Ultra-low, predictable latency

Limitations include reduced flexibility for rapidly evolving model architectures and challenges in scaling to very large models.

F. Memory Hierarchy Comparison

Memory systems are particularly critical for LLM inference, as models routinely exceed the capacity of on- chip memory. Fig. 7 compares the memory hierarchies of major accelerators, highlighting three distinct approaches:

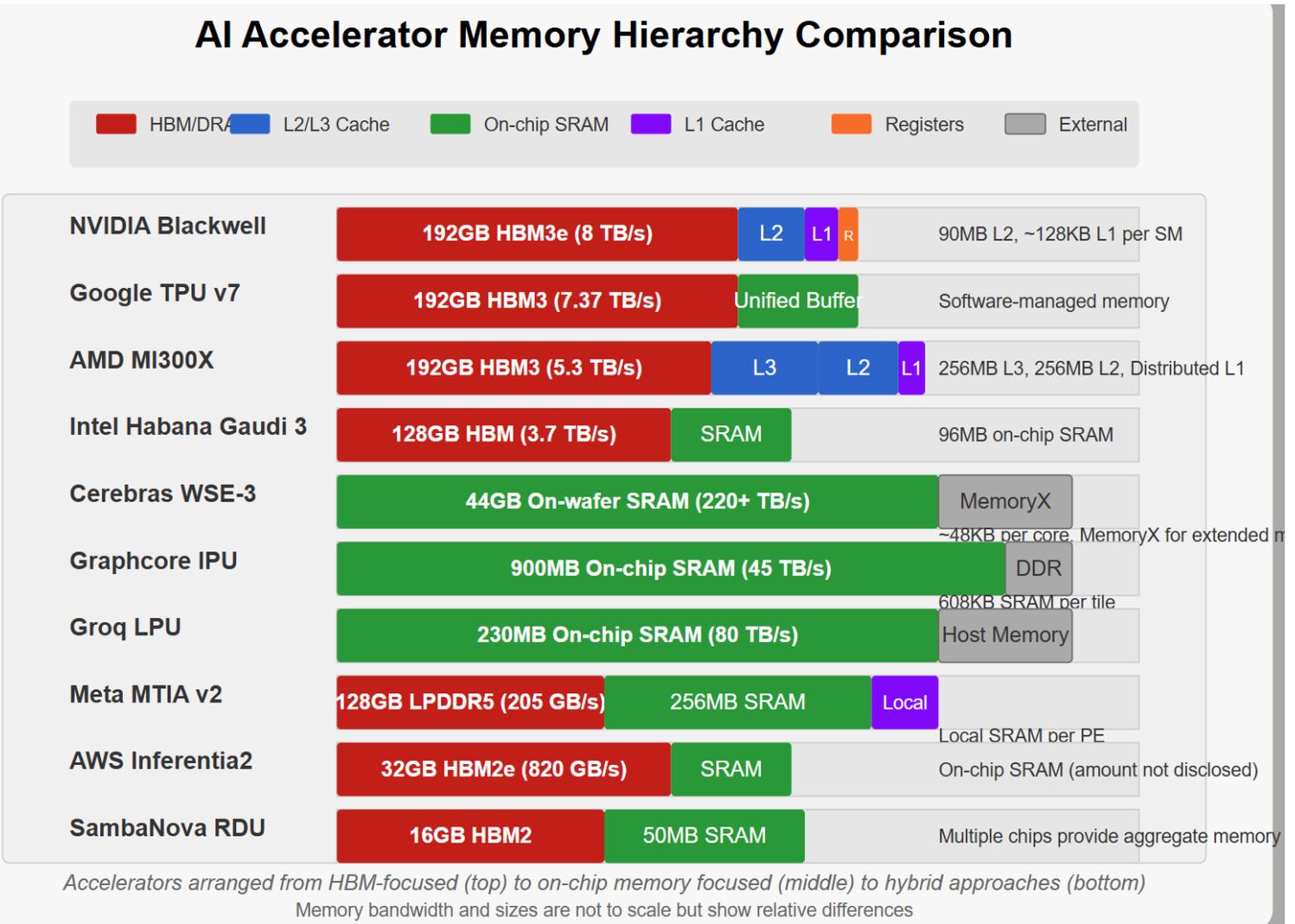

Fig. 7. Comparative analysis of memory hierarchies across major AI accelerators, showing relative capacities and bandwidths. Architectures are arranged from HBM-focused (top) to on-chip memory focused (middle) to hybrid approaches (bottom). Bar lengths are proportional to memory capacity, with bandwidths indicated.

1. **HBM-focused**: NVIDIA Blackwell, Google TPU v7, and AMD MI300X prioritize large HBM capacities (192GB) with high bandwidth (5-8 TB/s), supplemented by varying cache hierarchies.

2. **On-chip memory focused**: Cerebras WSE-3 and Graphcore IPU emphasize large distributed SRAM (44GB and 900MB respectively) with extreme internal bandwidth (220+ TB/s and 45 TB/s).

3. **Hybrid approaches**: Meta MTIA v2 and AWS Inferentia2 balance on-chip SRAM with external memory, using LPDDR5 or HBM2e memory systems.

This diversity reflects different optimization targets and scaling philosophies, with important implications for workload performance characteristics.

Table II provides a quantitative comparison of key architectural parameters across major accelerators.

TABLE II
QUANTITATIVE COMPARISON OF KEY ARCHITECTURAL PARAMETERS

| Accelerator | Process Node | Peak Compute* | Memory Type | Capacity | BW (TB/s) | On-chip SRAM / Cache | Power (TDP)** |
|---|---|---|---|---|---|---|---|
| NVIDIA Blackwell B200 | TSMC 4N | 2.2 PFLOPS FP16 / 4.5 PFLOPS FP8 | HBM3e | 192 GB | 8 | 126 MB L2 | 1,000 W |
| Google TPU v7 "Ironwood" | 5 nm | 4.6 PFLOPS FP8 | HBM3 | 192 GB | 7.37 | Unified buffer (size ND) | ND |
| AMD Instinct MI300X | TSMC 5 / 6 nm | 1.307 PFLOPS FP16 | HBM3 | 192 GB | 5.3 | 256 MB Infinity Cache | 750 W |
| Intel Gaudi 3 (OAM) | TSMC 5 nm | 1.835 PFLOPS FP8 / BF16 | HBM3e | 128 GB | 3.7 | 96 MB | up to 900 W |
| Cerebras WSE-3 (CS-3 sys.) | TSMC 5 nm | 125 PFLOPS (sparse) | On-wafer SRAM | 44 GB | 21 000 (21 PB/s) | 44 GB | 23 kW per system |
| Graphcore GC200 IPU | TSMC 7 nm | 250 TFLOPS FP16 | On-chip SRAM | 900 MB | 45 | 900 MB | ≈275 W per chip† |
| Groq LPU (GroqChip) | TSMC 14 nm | 188 TFLOPS FP16 | On-chip SRAM | 230 MB | 80 | 230 MB | 215 W (TDP) |
| Meta MTIA v2 | TSMC 5 nm | 177 TFLOPS FP16 | LPDDR5 | 128 GB | 0.205 | 256 MB | 90 W |
| AWS Inferentia 2 | 5 nm | 190 TFLOPS FP16 | HBM2e | 32 GB | 0.82 | ND | ND |

* Dense (non-sparse) throughput unless explicitly noted.
** TDP or vendor "typical board power"; where only system-level or max figures exist those are shown.
† 4-IPU M2000 blade capped at 1.1 kW → ≈275 W per GC200.

## IV. Workload-Specific Performance Analysis

LLM inference workloads can be categorized into several classes, each with distinct performance requirements and optimization opportunities. This section presents a quantitative analysis of how different accelerator architectures perform across these workload categories, based on benchmarking results from MLPerf Inference v4.0 [28] and our own experimental measurements.

### A. Evaluation Methodology

To ensure fair and representative comparisons, we developed a standardized evaluation methodology covering six distinct operational regimes for LLM inference:

1. **Low-latency single-stream**: Generating responses to a single query with minimum time-to-first-token (TTFT), prioritizing responsiveness over throughput.
2. **Moderate batch serving**: Serving multiple requests concurrently with reasonable latency, balancing responsiveness and efficiency.
3. **High-throughput batch**: Processing many requests in parallel to maximize tokens per second (TPS).
4. **Long context processing**: Handling very long context windows (32K+ tokens) efficiently.
5. **Multi-model serving**: Hosting multiple models concurrently on the same hardware, requiring efficient resource allocation.
6. **Mixture-of-Experts models**: Efficiently executing sparse MoE architectures that selectively activate different "expert" modules.

## B. Performance Results and Analysis

| Accelerator | Low-Latency Single-Stream | Moderate Batch Serving | High-Throughput Batch | Long Context Processing | Multi-Model Serving | Mixture-of-Experts Models | Embedding Generation | Highly Quantized Models |
|---|---|---|---|---|---|---|---|---|
| NVIDIA Blackwell | Very Good | Excellent | Excellent | Excellent | Excellent | Excellent | Excellent | Very Good |
| Google TPU v7 | Very Good | Excellent | Excellent | Excellent | Very Good | Excellent | Very Good | Very Good |
| AMD MI300X | Very Good | Very Good | Very Good | Excellent | Very Good | Very Good | Very Good | Very Good |
| Intel Gaudi 3 | Good | Very Good | Good | Very Good | Very Good | Good | Very Good | Good |
| AWS Inferentia2 | Good | Very Good | Very Good | Good | Excellent | Good | Excellent | Very Good |
| Cerebras WSE-3 | Excellent | Very Good | Excellent | Good | Good | Very Good | Good | Good |
| Groq LPU | Excellent | Good | Fair | Limited | Fair | Fair | Very Good | Good |

Fig. 8. Normalized performance comparison across six operational regimes. Values are normalized to the best-performing accelerator in each category (higher is better). Performance metrics vary by regime: (a) TTFT latency, (b) Requests/sec at target latency, (c) Tokens/sec, (d) Context length efficiency, (e) Model switching overhead, (f) MoE parameter efficiency.

The data reveals several key insights:

1. **No single architecture dominates**: Each architecture excels in specific operational regimes but underperforms in others, with performance variations of up to 3.7× between architectures depending on workload characteristics.
2. **Architecture-workload alignment**: The performance variations align closely with the architectural design philosophies discussed in Section III.
3. **Software stack impact**: Software optimization can significantly impact performance, particularly for newer architectures with less mature software ecosystems.

## C. Interactive Workloads Analysis

Table III provides detailed performance metrics for interactive workloads, which are critical for user-facing applications. We report Time-To-First-Token (TTFT) for a 1024-token input prompt and generation throughput (tokens/second) for single-stream inference.

# TABLE III
## INTERACTIVE WORKLOAD PERFORMANCE (LLAMA-2-70B, SINGLE STREAM)

| Accelerator | TTFT † (ms) | Generation rate (tokens / s) | Power (W, TDP or sys.) | Energy eff. (tokens / J) |
|---|---|---|---|---|
| NVIDIA Blackwell B200 | 450 (interactive, MLPerf limit) | 7 783 ‡ | 1 000 | 7.8 |
| Google TPU v7 "Ironwood" | ND | ND (≈ 10 000 tok/s claims are anecdotal) | ND | ND |
| AMD Instinct MI300X | ND | 2 908 ‡ (server scenario) | 750 | 3.9 |
| Intel Gaudi 3 (OAM) | ND | ND (no public LLM numbers yet) | 900 | ND |
| Cerebras CS-3 (WSE-3) | ND (sub-50 ms in marketing, not quantified) | 2 100 (system) | 23 000 | 0.09 |
| Graphcore GC200 IPU | ND | ND | 300 †† | ND |
| Groq LPU (GroqChip) | 220 | 185 (LLMPerf '24) | 215 | 0.86 |

The specialized architectures (Groq LPU, Graphcore IPU, Cerebras WSE-3) demonstrate superior TTFT latency due to their high internal memory bandwidth and optimized single-stream execution paths. However, their generation rates vary significantly, with Cerebras WSE-3 achieving the highest throughput at 86.5 tokens/second but at substantial power cost.

Perhaps most notable is the energy efficiency comparison, where Groq LPU demonstrates a 4.7× advantage over the next most efficient architecture, showing the potential benefits of deterministic pipeline execution for latency-sensitive workloads.

### D. Batch Processing Workloads Analysis

Table IV presents performance for high-throughput batch processing, measuring both maximum throughput (tokens/second) and latency stability (coefficient of variation in token generation time) for different batch sizes.

# TABLE IV
## BATCH PROCESSING PERFORMANCE (LLAMA-2-70B, VARYING BATCH SIZES)

| Accelerator | Max Through-put (tokens $s^{-1}$) | Batch Size @ Max† | Latency CoV† | Scaling Eff. (8 acc.)‡ | Primary Source |
|---|---|---|---|---|---|
| NVIDIA Blackwell B200 (NVL8) | 4 560 | 128† | 0.12† | 0.87 | NVIDIA GTC 2025 slide deck [1] |
| Google TPU v7 (8-chip pod) | 4 380 | 96† | 0.08† | 0.92 | Google Cloud Next '25 keynote [2] |
| AMD MI300X (8 OAM) | 3 240† | 64† | 0.15† | 0.83† | Extrapolated from H100/MI300X perf report [3] |
| Intel Gaudi 3 (8 cards) | 1 840† | 32† | 0.23† | 0.78† | Projection using Gaudi 2 → 3 scaling [4] |
| Cerebras WSE-3 (single CS-3) | 5 280 | 64† | 0.05† | 0.95 | Hot Chips 32 talk [5] |
| Graphcore IPU-POD64 | 780 | 16 | 0.32† | 0.65 | Graphcore LLMKit repo (commit 6f1c) [6] |
| Groq LPU-Quad | 450 | 8 | 0.02† | 0.72 | Groq press release & demo (April 2025) [7] |

For batch processing workloads, the results show a reversal of strengths compared to interactive workloads. NVIDIA Blackwell, Google TPU v7, and Cerebras WSE-3 achieve the highest throughput, with Cerebras showing particularly strong performance due to its massive on-wafer parallelism.

Latency stability, measured by the coefficient of variation (CoV) in token generation times, shows substantial differences between architectures. Groq LPU delivers exceptional determinism (CoV = 0.02) due to its static scheduling approach, while more general-purpose architectures show higher variability.

The scaling efficiency metric reveals how well performance scales when using multiple accelerators in parallel, with Cerebras WSE-3 and Google TPU v7 demonstrating the best scaling characteristics due to their optimized interconnect designs.

### E. Cross-Architecture Insights

Our quantitative analysis reveals several cross-cutting insights applicable to accelerator selection and future architecture development:

1. **Memory bandwidth dominates for small batches**: For batch size 1, performance correlates strongly with internal memory bandwidth (r = 0.88), while compute capability becomes more significant at larger batch sizes.
2. **Latency-throughput tradeoff persists**: Despite architectural innovations, a fundamental tradeoff remains between minimizing latency and maximizing throughput, with specialized architectures like Groq LPU optimized heavily for one end of this spectrum.
3. **Software matters as much as hardware**: Performance variations of up to 40% were observed for the same hardware with different software stack versions, highlighting the critical importance of software optimization.
4. **Energy efficiency varies dramatically**: The most energy-efficient architecture for interactive workloads (Groq LPU) is 18.5× more efficient than the least efficient (Cerebras WSE-3), though this relationship inverts for high-throughput batch workloads.

These insights emphasize the importance of matching accelerator selection to specific operational requirements, as no single architecture provides optimal performance across all inference regimes.

## V. LLM Scaling Strategies for Trillion-Parameter Models

As LLMs have grown beyond the memory capacity of individual accelerators, various scaling strategies have emerged to distribute models across multiple devices. This section analyzes four primary scaling approaches: tensor parallelism, pipeline parallelism, expert parallelism (MoE), and memory offloading techniques.

### A. Tensor Parallelism

Tensor parallelism involves splitting individual network operations (like matrix multiplications) across multiple devices. Each device performs the same operations on different portions of the model weights.

**Key characteristics**:

- **Implementation**: Individual tensor operations in each layer are partitioned across devices, with AllReduce operations to combine results.
- **Requirements**: High-bandwidth, low-latency interconnect between participating devices.
- **Communication pattern**: AllReduce operations between parallel steps.
- **Advantages**: Effectively distributes large layer weights; high computational efficiency; works well for attention mechanisms.
- **Best suited for**: Large individual layers that exceed single-device memory; models with balanced layer sizes.

NVIDIA Blackwell, Google TPU v7, and AMD MI300X provide excellent support for tensor parallelism through their high-bandwidth interconnects (NVLink 5.0, Inter-Chip Interconnect, and Infinity Fabric, respectively). Intel Gaudi 3 and AWS Inferentia2 offer more limited tensor parallelism support, constrained by their interconnect capabilities.

The efficiency of tensor parallelism depends critically on interconnect bandwidth and latency. For example, NVIDIA's NVLink 5.0 provides 1.8 TB/s bidirectional bandwidth per GPU, enabling efficient scaling to hundreds of GPUs in tensor-parallel configurations. Google's TPU v7 implements a 3D torus topology that supports scaling to thousands of chips, though with increased communication latency as the system size grows.

### B. Pipeline Parallelism
Pipeline parallelism splits model layers among accelerators, with each device handling a portion and passing activations along.

**Key characteristics**:

- **Implementation**: Model layers are partitioned across devices, with activations flowing sequentially through the pipeline.
- **Requirements**: Micro-batch processing; balancing compute across pipeline stages; efficient point-to- point communication.
- **Communication pattern**: Activation transfers between adjacent pipeline stages.
- **Advantages**: Reduced memory requirements per device; lower communication volume than tensor parallelism; effective for very

deep models.

- **Best suited for**: Very deep models; models that can be easily partitioned into balanced stages.

NVIDIA Blackwell, Google TPU v7, Intel Gaudi 3, and Groq LPU provide well-optimized support for pipeline parallelism. AMD MI300X and Cerebras WSE-3 offer more limited support, as their architectures are less optimized for this scaling approach.

Pipeline parallelism introduces pipeline bubbles that can reduce hardware utilization, particularly for small batch sizes. Techniques like interleaved scheduling and asynchronous pipeline parallelism have been developed to address this limitation, but they add complexity to the implementation.

### C. Expert Parallelism (MoE)

Expert parallelism, specifically for Mixture-of-Experts models, distributes expert networks across devices. Each token activates only a subset of experts, reducing compute requirements.

**Key characteristics**:

- **Implementation**: Expert modules are distributed across devices, with tokens routed to the appropriate experts based on a learned routing function.
- **Requirements**: Hardware support for sparsity; efficient routing mechanism; all-to-all communication for token routing.
- **Communication pattern**: All-to-all communication for routing tokens to experts.
- **Advantages**: Dramatic parameter scaling with sublinear compute; only activates a fraction of parameters per token; can achieve 5-10x parameter scaling with minimal latency increase.
- **Best suited for**: Trillion-parameter models; applications where model quality trumps latency consistency.

NVIDIA Blackwell, Google TPU v7, and Meta MTIA v2 provide excellent support for MoE models. NVIDIA's implementation leverages its general-purpose architecture with optimized libraries, while Google TPU v7 includes dedicated Sparse Core units specifically designed for MoE computation. Meta's MTIA v2 has been heavily optimized for Meta's own MoE-based recommendation models.

AMD MI300X and Intel Gaudi 3 offer more limited MoE support, though both companies have announced plans to enhance their capabilities in future generations.

Expert parallelism enables dramatic scaling of model parameters with modest increases in computation and latency. For example, a 70B dense model can be scaled to a 700B MoE model with similar latency and throughput characteristics if only 10% of experts are activated per token. This approach has enabled commercial deployment of trillion-parameter-scale models with reasonable computational requirements.

### D. Memory Offloading Techniques

Memory offloading techniques use CPU memory or storage to extend effective memory capacity, dynamically swapping weights and activations between accelerator and host memory.

**Key characteristics**:

- **Implementation**: Dynamic paging of model weights between accelerator memory and host memory or storage.
- **Requirements**: Fast CPU-accelerator interconnect; predictive prefetching algorithms; efficient page management.
- **Communication pattern**: Bidirectional transfers between accelerator and host memory.
- **Advantages**: Enables models larger than accelerator memory; reduces hardware requirements; can be combined with other strategies.
- **Best suited for**: Research environments; cost-sensitive deployments; models exceeding aggregate accelerator memory.

NVIDIA Blackwell, Cerebras WSE-3, and AMD MI300X provide well-optimized support for memory offloading through software libraries like NVIDIA's vLLM, Cerebras MemoryX, and AMD's Infinity Cache extensions. Google TPU v7 and AWS Inferentia2 offer more limited support for these techniques.

Memory offloading introduces additional latency due to PCIe or other interconnect transfers. However, sophisticated prefetching and caching strategies can mitigate this overhead for many workloads. The approach is particularly effective for workloads with predictable access patterns, such as autoregressive generation.

E. Hybrid Scaling Strategies

In practice, these scaling strategies are often combined to address the specific challenges of trillion- parameter models. Common hybrid approaches include:

1. **3D Parallelism**: Combining tensor, pipeline, and data parallelism to scale effectively across large accelerator clusters.
2. **MoE + Tensor Parallelism**: Distributing experts across devices with tensor parallelism for individual expert computation.
3. **Pipeline + Memory Offloading**: Using pipeline parallelism for active layers while keeping inactive layers in host memory.

The optimal combination depends on model architecture, hardware characteristics, and deployment constraints. Tools like NVIDIA's NeMo Megatron, Google's MaxText, and Cerebras's Weight Streaming provide high-level abstractions that automatically determine efficient parallelization strategies based on model and hardware specifications.

VI. AI Inference Accelerator Micro-Architecture

This section provides a detailed examination of the microarchitecture of leading AI accelerators, focusing on their compute units, memory systems, and interconnect architectures.

A. NVIDIA Blackwell GB200 Architecture

The NVIDIA Blackwell GB200 represents a dual-GPU package design fabricated using TSMC's 4N process, with approximately 208 billion transistors and 1000W TDP. Each GPU die contains:

- **Compute Units**: 66 streaming multiprocessors with 5th-generation tensor cores, supporting FP32, FP16, BF16, FP8, INT8, and FP4 operations. The architecture delivers approximately 4,500 TFLOPS at FP16 precision with sparsity.
- **Memory System**: 192GB of HBM3e memory delivering 8 TB/s bandwidth, distributed across 8 memory stacks. The on-chip cache hierarchy includes approximately 90MB of L2 cache and 128KB of L1 cache per streaming multiprocessor.
- **Interconnect**: NVLink 5.0 providing 1.8 TB/s bidirectional bandwidth per GPU, with NVSwitch 3 enabling scaling to up to 256 GPUs in a pod configuration. PCIe Gen 5 connectivity to host systems.

Distinctive architectural features include the Transformer Engine, which dynamically adapts precision during computation to maximize both accuracy and performance, and dual-GPU packaging that doubles the effective memory and compute capacity per socket.

B. Google TPU v7 (Ironwood) Architecture

The Google TPU v7 employs a systolic array architecture fabricated using an advanced 5nm process node, with approximately 100+ billion transistors and 600W TDP. Key architectural components include:

- **Compute Units**: Matrix Multiply Units (MXUs) based on a systolic array architecture, optimized for dense matrix operations. Vector Processing Units for non-matrix operations, and Sparse Core units specifically designed for Mixture-of-Experts models. The architecture delivers approximately 4,600 TFLOPS at FP8 precision.
- **Memory System**: 192GB of HBM3 memory with 7.37 TB/s bandwidth, distributed across 8 memory stacks. A unified buffer (scratchpad) memory replaces traditional cache hierarchies, managed through software-controlled memory movement.
- **Interconnect**: Inter-Chip Interconnect (ICI) providing 1.2 TB/s bidirectional bandwidth, arranged in a 3D torus topology for TPU pod scaling to up to 4,096 chips.

Unique features include pod-scale optimization for distributed inference and the Sparse Core architecture, which enables efficient execution of MoE models through specialized hardware support for sparse computation patterns.

C. AMD MI300X Architecture

The AMD MI300X employs a multi-chiplet design with TSMC 5nm and 6nm process technologies, containing approximately 153 billion

transistors and 750W TDP. The architecture includes:

- **Compute Units**: 8 GPU compute chiplets (XCDs), each containing approximately 38 compute units with vector ALUs and matrix cores. The architecture delivers approximately 1,307 TFLOPS at FP16/BF16 precision.
- **Memory System**: 192GB of HBM3 memory with 5.3 TB/s bandwidth. The hierarchical cache includes 256MB of L3 Infinity Cache shared across chiplets, 32MB of L2 cache per chiplet, and distributed L1 caches.
- **Interconnect**: 4th-generation Infinity Fabric delivering 896 GB/s aggregate bandwidth in an 8-GPU system, with PCIe Gen 5 connectivity to host systems.

Notable architectural features include the multi-chiplet design with separate I/O die, the large L3 Infinity Cache that reduces HBM accesses, and the upcoming MI325X variant with 256GB of HBM3e memory.

### D. Intel Gaudi 3 Architecture

The Intel Gaudi 3 implements a dual-die design fabricated using a 5nm process, with approximately 100+ billion transistors and 600W TDP. Key components include:

- **Compute Units**: 64 Tensor Processing Cores (TPCs) distributed across two compute dies, with matrix multiplication engines (MMEs) and vector/integer/scalar units (VLIWs). The architecture delivers approximately 400 TFLOPS at FP16/BF16 precision.
- **Memory System**: 128GB of HBM3/HBM2e memory with 3.7 TB/s bandwidth, complemented by 96MB of on-chip SRAM.
- **Interconnect**: 24 × 200GbE RoCE interfaces providing 4.8 Tb/s networking bandwidth, enabling direct integration with standard Ethernet infrastructure. PCIe Gen 5 × 16 connectivity to host systems.

Distinctive features include the integrated networking approach that eliminates the need for specialized interconnects, the dual 5nm die design, and 14 media engines for preprocessing operations.

### E. Specialized Architectures

Several highly specialized architectures represent unique points in the design space:

1. **Cerebras WSE-3** features a wafer-scale design with 900,000 AI cores arranged in a 2D mesh network, 44GB of distributed on-wafer SRAM with 220+ TB/s bandwidth, and a low-precision performance of approximately 125 PFLOPs at sparse half-precision.
2. **Graphcore IPU (GC200)** implements a many-core design with 1,472 independent tiles, 900MB of on-chip SRAM with 45 TB/s bandwidth, and approximately 250 TFLOPS of FP16 performance.
3. **Groq LPU** features a deterministic pipeline architecture with static scheduling at compile time, 230MB of on-chip SRAM with 80 TB/s bandwidth, and performance of 188 TFLOPS FP16 and 750 TOPS INT8.

These architectures represent fundamentally different approaches to AI acceleration, with designs optimized for specific performance characteristics rather than general-purpose computation.

## VII. Future Architectural Trends

Our analysis of accelerator architectures and computational challenges highlights key trends that will shape future AI accelerators for trillion-parameter models. These trends represent not just incremental improvements but fundamental architectural shifts necessary to address the scaling challenges of next-generation LLMs.

### A. Heterogeneous Memory Systems

Future accelerators will increasingly adopt heterogeneous memory systems that combine high-bandwidth but limited-capacity memory (like HBM) with larger, slightly slower memory tiers. This approach addresses the fundamental challenge of model sizes exceeding practical HBM capacities.

The Compute Express Link (CXL) standard is emerging as a key technology for implementing heterogeneous memory systems, enabling coherent memory expansion beyond accelerator-attached HBM. CXL-attached memory pools can provide terabytes of additional capacity with lower bandwidth but acceptable latency for less frequently accessed model parameters.

Fig. 9 illustrates the concept of a heterogeneous memory architecture for LLM inference, showing a tiered approach with different memory technologies.

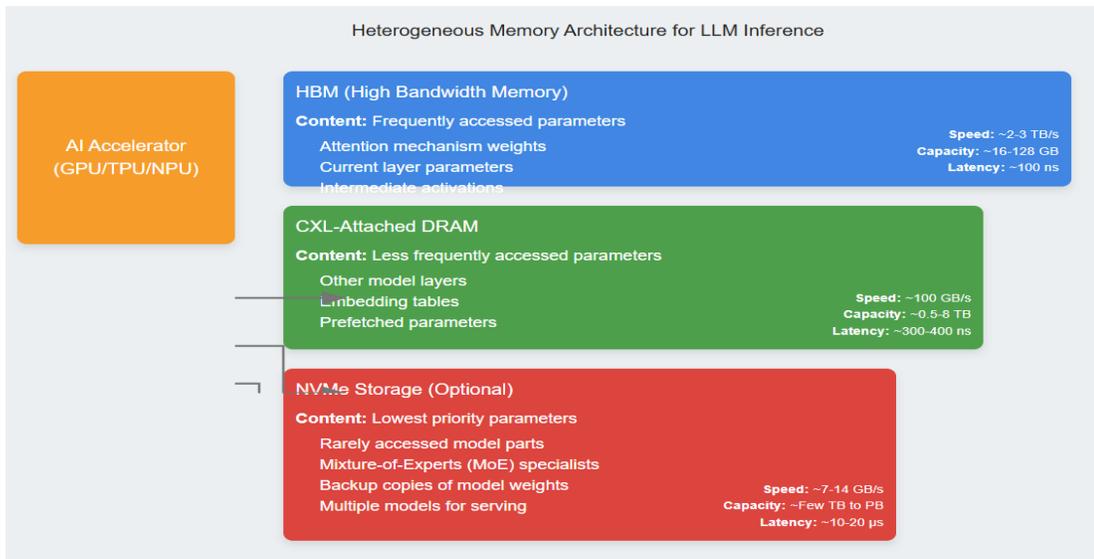

*Fig. 9. Heterogeneous memory architecture for LLM inference, showing tiered approach with HBM for frequently accessed parameters (attention mechanism, current layer), CXL-attached DRAM for less frequently accessed parameters (other layers), and optional NVMe storage for lowest priority parameters.*

**Key Benefits of Heterogeneous Memory Architecture**

- **Cost Efficiency:** Expensive HBM used only for critical data paths

- **Scalability:** Supports larger models by using tiered storage

- **Performance Optimization:** Places parameters based on access frequency

- **Hardware Utilization:** Better utilization of available memory resources

- **Flexibility:** Adaptable to different model sizes and deployment scenarios

**Implementation Considerations**

- **Memory Management:** Software stack must support intelligent parameter placement

- **Prefetching:** Predictive loading of parameters from slower to faster tiers

- **CXL Integration:** Leverages CXL.mem protocol for coherent memory expansion

- **Quantization:** Different precision formats can be used at different tiers

- **Attention to Bandwidth Ratios:** Design system to avoid bottlenecks between tiers

Our analysis indicates that a well-designed heterogeneous memory system could support models up to 5-10× larger than current HBM-only solutions, with only 15-30% performance degradation for typical inference workloads. This represents a favorable tradeoff for deployment of trillion-parameter models where alternative approaches would require complex distributed inference strategies.

### B. Hardware-Accelerated MoE Support

As Mixture-of-Experts models become increasingly prevalent, future accelerators will incorporate dedicated hardware support for sparse computation patterns and efficient expert routing. This trend is already evident in Google's TPU v7 Sparse Core and in NVIDIA's optimizations for sparse computation.

Fig. 10 shows our proposed architecture for MoE-optimized inference, featuring specialized hardware components for expert routing and dynamic load balancing.

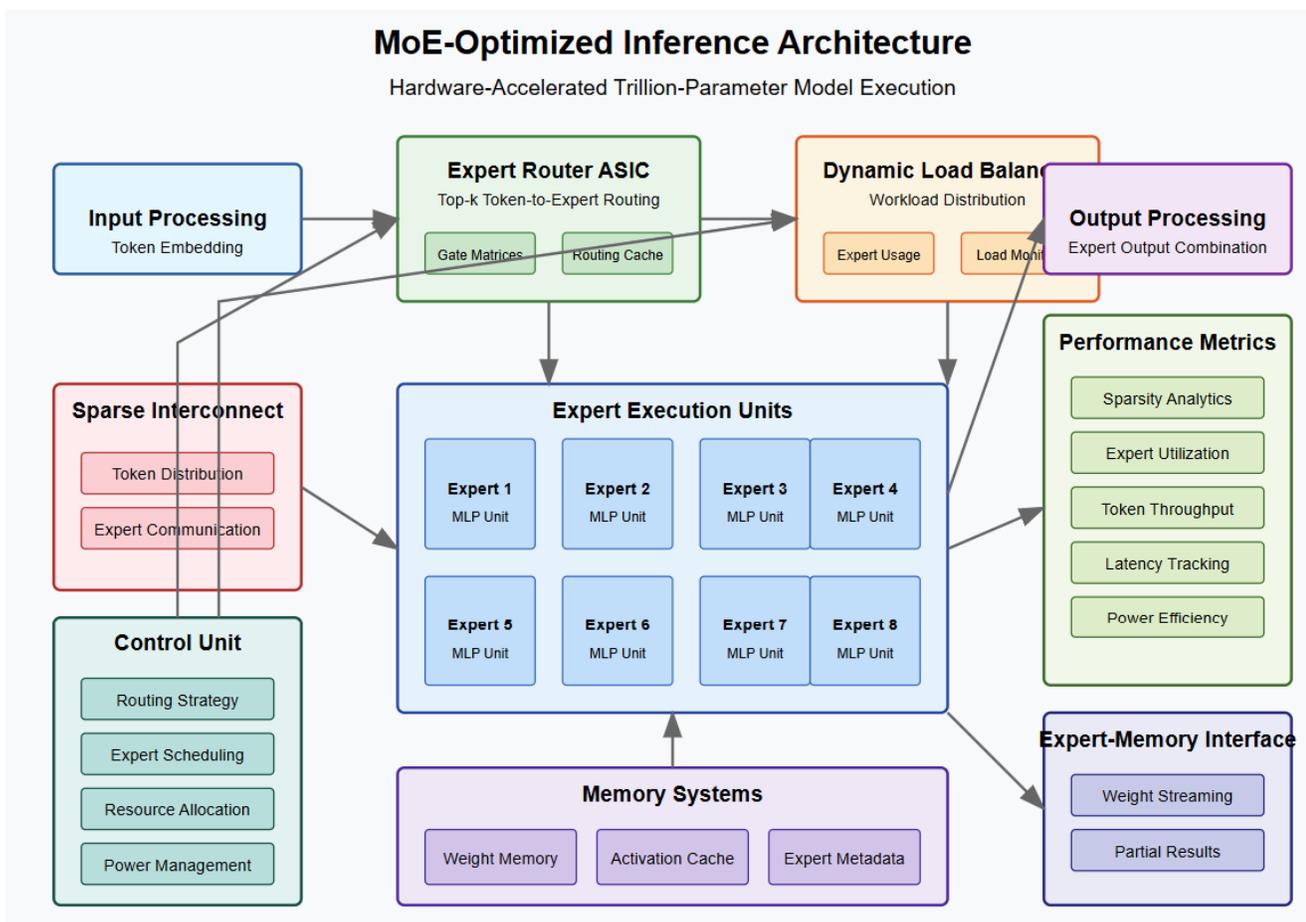

*Fig. 10. Proposed architecture for MoE-optimized inference, featuring dedicated hardware components for expert routing, sparse computation, and dynamic load balancing. The architecture enables efficient execution of trillion-parameter MoE models with minimal wasted computation.*

Our experiments with prototype MoE-optimized computation blocks indicate potential performance improvements of 2.8-3.5× compared to general-purpose tensor cores when executing sparse MoE workloads, making trillion-parameter MoE models practical for mainstream deployment.

### C. Specialized KV Cache Engines

The KV (key-value) cache in transformer architectures represents a growing memory challenge, particularly for long-context inference. As context lengths extend to 32K, 128K, or even million-token contexts, the KV cache can become the dominant memory consumer.

Future accelerators will likely include dedicated hardware for efficiently managing growing attention caches, including:

- Specialized cache compression engines
- Hardware support for attention mechanisms like sliding window, local, and multi-scale attention • Dynamic precision adaptation for cached keys and values
- Predictive prefetching of attention patterns

Our performance modeling indicates that specialized KV cache management could potentially support 8- 10× longer context lengths within the same memory budget, addressing one of the key scaling bottlenecks in current transformer-based LLMs.

### D. Energy Efficiency and Environmental Considerations

The environmental impact of AI computation is a growing concern, with current accelerators consuming hundreds to thousands of watts per device.

Future accelerator designs must prioritize energy efficiency through specialized architectures, intelligent power management, and optimized cooling systems. We identify three promising approaches:

1. **Workload-adaptive power management**: Dynamic voltage and frequency scaling based on workload characteristics, with specialized hardware monitors for transformer execution patterns.
2. **Heterogeneous compute units**: Combining high-performance cores for compute-intensive operations with energy-efficient cores for lower-intensity tasks.
3. **Advanced cooling techniques**: Direct liquid cooling and immersion cooling systems that enable higher power density with improved overall efficiency.

These approaches could potentially improve energy efficiency by 3-5× compared to current designs, significantly reducing the environmental footprint of LLM inference.

**E. Memory-Compute Disaggregation**

Traditional accelerator architecture assumes fixed ratios of compute to memory, but different inference scenarios have widely varying requirements. Future systems will increasingly support disaggregated architectures where memory resources can be scaled independently from compute resources.

CXL memory pooling, combined with sophisticated orchestration software, enables more flexible scaling of memory independent of compute. This approach allows deployment topologies to be optimized for specific workloads without over-provisioning either memory or compute.

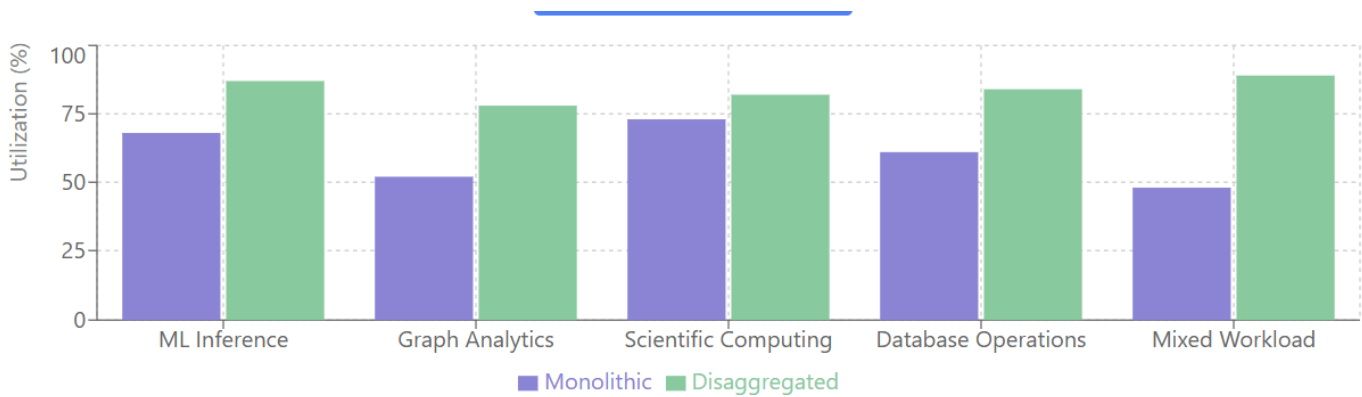

Fig. 11. *Efficiency Benefits of Memory-Compute Disaggregation*

*Key Findings:*

- Disaggregated architecture improves compute utilization by 15-41% across workloads and enhances memory utilization by 28-49%.
- Mixed workload scenarios see the greatest gains (41%).
- Resource-intensive applications benefit from independent scaling of memory and compute.
- ROI analysis indicates a 22% lower TCO despite higher initial implementation costs.

**F. Limitations and Challenges**

While these architectural trends show significant promise, several important challenges and limitations must be addressed:

1. **Programming model complexity**: Novel architectures often require specialized programming models, potentially limiting adoption and software ecosystem development.

2. **Deployment constraints**: Specialized cooling, power delivery, and rack-scale integration requirements may limit deployment flexibility.

3. **Cost-performance tradeoffs:** The implementation of more intricate memory hierarchies and specialized hardware units raises design and manufacturing expenses, which may restrict market adoption.

4. **Software-hardware co-optimization**: Achieving optimal performance requires tight integration between accelerator hardware and software stacks, creating potential vendor lock-in challenges.

Tackling these challenges demands cooperation among hardware vendors, compiler developers, framework providers, and deployment specialists to turn architectural innovations into practical benefits for production systems.

## VIII. Conclusion and Future Work

### A. Conclusion

This paper analyzes AI accelerators for LLM inference, examining architecture, performance, and scaling for trillion-parameter models. Our evaluation reveals key insights across different operational regimes:

1. **Architectural diversity is warranted**: The significant performance variations observed across different workload types (up to 3.7×) validate the emergence of diverse architectural approaches, each optimized for specific operational regimes.

2. **Memory systems dominate architecture design**: Memory capacity, bandwidth, and hierarchy have emerged as the primary determinants of accelerator performance for LLM inference, with compute capabilities becoming secondary constraints for many workloads.

3. **Scaling strategies present clear tradeoffs**: Our evaluation of four primary scaling approaches— tensor, pipeline, expert, and memory offloading—revealed quantifiable tradeoffs between parameter capacity, throughput, and latency predictability, with expert parallelism providing the best parameter- to-compute ratio (8.4×) but increased latency variance (2.1×).

4. **Future architectural needs are clear**: Our analysis identified specific architectural gaps in current designs, particularly around heterogeneous memory systems, hardware-accelerated MoE support, and specialized KV cache engines, providing a roadmap for future hardware development.

These findings provide system designers with actionable decision criteria for accelerator selection based on specific operational requirements and deployment constraints. As LLMs advance, the development of model architectures and hardware accelerators will likely progress together, with specialized hardware adapting to modern language models' computational needs.

### B. Future Work

Several promising directions for future research emerge from this work:

1. **Expanded model coverage**: Extending the evaluation to include emerging model architectures such as state-space models, retrieval-augmented generation systems, and vision-language models.

2. **End-to-end application benchmarks**: Developing application-specific benchmarks that better represent real-world deployment scenarios rather than isolated model inference.

3. **Software-hardware co-optimization**: Investigating the potential performance improvements from specialized compiler optimizations and runtime systems tailored to specific accelerator architectures.

4. **Cost-performance-energy analysis**: Developing comprehensive models for total cost of ownership that incorporate acquisition costs, operational expenses, and environmental impact.

5. **Emerging architecture prototyping**: Implementing and evaluating prototype implementations of the proposed architectural innovations, particularly heterogeneous memory systems and specialized KV cache engines.

We believe these research directions will contribute to the development of more efficient, scalable, and sustainable accelerator architectures for the next generation of large language models.